\documentclass[pdflatex, referee, sn-nature]{sn-jnl}% Style for submissions to Nature Portfolio journals
\usepackage{graphicx}%
\usepackage{multirow}%
\usepackage{amsmath,amssymb,amsfonts}%
\usepackage{amsthm}%
\usepackage{mathrsfs}%
\usepackage[title]{appendix}%
\usepackage{xcolor}%
\usepackage{textcomp}%
\usepackage{manyfoot}%
\usepackage{booktabs}%
\usepackage{algorithm}%
\usepackage{algorithmicx}%
\usepackage{algpseudocode}%
\usepackage{listings}%

\usepackage[utf8]{inputenc}
\usepackage{booktabs}
\usepackage{caption}
\usepackage{geometry}
\usepackage{setspace}
\usepackage{etoolbox}
\theoremstyle{thmstyleone}%
\theoremstyle{thmstyletwo}%

\theoremstyle{thmstylethree}%

\begin{document}

\title{\textbf{Giant Nonlinear Photon-Drag Currents in Moiré Bilayers}}


\author[1]{\fnm{Zhuocheng} \sur{Lu}}
\equalcont{These authors contributed equally to this work.}

\author[2]{\fnm{Zhuang} \sur{Qian}}
\equalcont{These authors contributed equally to this work.}

\author[1]{\fnm{Zhichao} \sur{Guo}}

\author[1]{\fnm{Likun} \sur{Shi}}

\author[2]{\fnm{Shi} \sur{Liu}}

\author*[1]{\fnm{Hua} \sur{Wang}}\email{daodaohw@zju.edu.cn}

\author*[1]{\fnm{Kai} \sur{Chang}}\email{kchang@zju.edu.cn}

\affil*[1]{\orgdiv{Center for Quantum Matter, School of Physics}, \orgname{Zhejiang University}, \orgaddress{ \city{Hangzhou}, \postcode{310058}, \state{Zhejiang}, \country{China}}}

\affil[2]{\orgdiv{Department of Physics, School of Science}, \orgname{Westlake University}, \orgaddress{ \city{Hangzhou}, \postcode{310030}, \state{Zhejiang}, \country{China}}}


\abstract{
The bulk photovoltaic effect provides a fundamental pathway for direct light-to-current conversion in quantum materials. However, these nonlinear currents are often strictly constrained or forbidden by crystal symmetries, hindering their exploration in a broader range of materials. While the nonlinear photon-drag effect leverages finite photon momentum to circumvent these constraints, its investigation has been largely confined to toy models, lacking a robust numerical framework for realistic materials. Here, we develop a unified microscopic theory of nonlinear photon-drag currents formulated within a geometric-loop framework, providing both a transparent quantum-geometric interpretation and numerical tractability. Applying this formalism to twisted bilayer graphene (TBG), we demonstrate that a finite, in-plane photon momentum can trigger massive nonlinear responses, rivaling the giant photovoltaic currents reported in typical 2D materials. These currents exhibit high tunability via photon wavevector, twist angle, and light polarization. Our work not only provides a generalized framework for momentum-dependent light-matter interactions but also establishes the nonlinear photon-drag effect as a potent mechanism for unlocking unprecedented optoelectronic functionalities beyond the limitations of the conventional bulk photovoltaic effect.
}

\maketitle

\clearpage

\section{Introduction}\label{sec1}

The generation of dc photocurrents via the bulk photovoltaic effect represents a fundamental mechanism for direct light-to-current conversion in quantum materials \cite{hosur2011, morimoto2016, dejuan2017, rangel2017, wang2020, ahn2020, ahn2022, chaudhary2022, resta2024, sivianes2025, avdoshkin2025}. However, these nonlinear currents are strictly constrained or entirely forbidden by crystal symmetries (such as mirror or spatial inversion), hindering their exploration in a broader range of material systems. Although this restriction can be bypassed by accounting for finite momentum transfer from photons to carriers, the resulting photon-drag effect is typically negligible in conventional settings due to the vanishingly small momentum of free-space optical photons. The situation changes when the light field is engineered to enhance momentum transfer. Structured beams with spatially varying phase and amplitude, such as tightly focused Gaussian beams, locally break spatial symmetries and effectively supply large momenta through spatial dispersion, enabling measurable photovoltaic signals \cite{ji2019}. Polaritonic confinement provides a complementary route: subwavelength surface plasmons and related polaritons compress optical wavelengths by tens to hundreds of times, generating large wavevectors and intense near-field gradients that can amplify the photon-drag response by orders of magnitude \cite{lezec2002, basov2016, kurman2018}. These developments demonstrate that enhanced photon-drag can successfully circumvent crystal symmetry constraints and unlock entirely new regimes of nonlinear optoelectronic response.

Despite early progress establishing photon-drag injection and shift currents as the dominant contributions to these responses, quantitative modeling has been largely confined to simplified toy-model analyses \cite{shi2021, xie2025}. A key bottleneck is computing the nonlinear photon-drag shift current in realistic electronic structures: physically transparent formulations often rely on gauge-dependent quantities that hinder stable numerical evaluation, whereas gauge-invariant sum-rule expressions obscure the underlying quantum geometry by requiring computationally heavy summations over all remote bands. Compounding these theoretical challenges, photon-drag phenomena in emerging moiré quantum materials remain largely unexplored. Among them, twisted bilayer graphene (TBG) stands out as a highly tunable platform hosting flat bands and a rich landscape of correlated and topological phases \cite{cao2018, cao2018a, lu2019, serlin2020, xie2021, stepanov2021, zhang2019, abouelkomsan2020}. While pristine TBG prohibits the conventional in-plane bulk photovoltaic effect due to its intrinsic $C_{2z}$ symmetry, its enlarged moiré supercell radically compresses the Brillouin zone. Consequently, the kinematic significance of a finite photon wavevector is profoundly amplified, positioning TBG as a premier system for realizing massive nonlinear photon-drag responses.

In this work, we develop a geometric-loop formalism for nonlinear photon-drag currents that unifies the injection and shift contributions within a single, gauge-invariant, and transparent framework, enabling both conceptual clarity and robust numerical implementation. We derive the explicit symmetry constraints on the nonlinear photon-drag photoconductivity tensor and apply this formalism to TBG using an exact continuum model calibrated with ab initio precision. Our calculations reveal substantial photon-drag responses that match the magnitude of the giant bulk photovoltaic effects predicted in prominent 2D semiconductors. We further characterize the scaling behavior of these photocurrents with the photon wavevector and map out their sensitive dependence on the twist angle. Finally, we analyze TBG aligned with a hexagonal boron nitride (hBN) substrate, where the substrate-induced sublattice asymmetry lifts the $C_{2z}$ symmetry constraint, thereby activating the conventional in-plane bulk photovoltaic effect. The clear disparities in magnitude and $\boldsymbol{k}$-space textures allow us to unambiguously distinguish between the conventional and photon-drag mechanisms.

\begin{figure}[htbp]
    \centering
    \includegraphics[width=0.9\columnwidth]{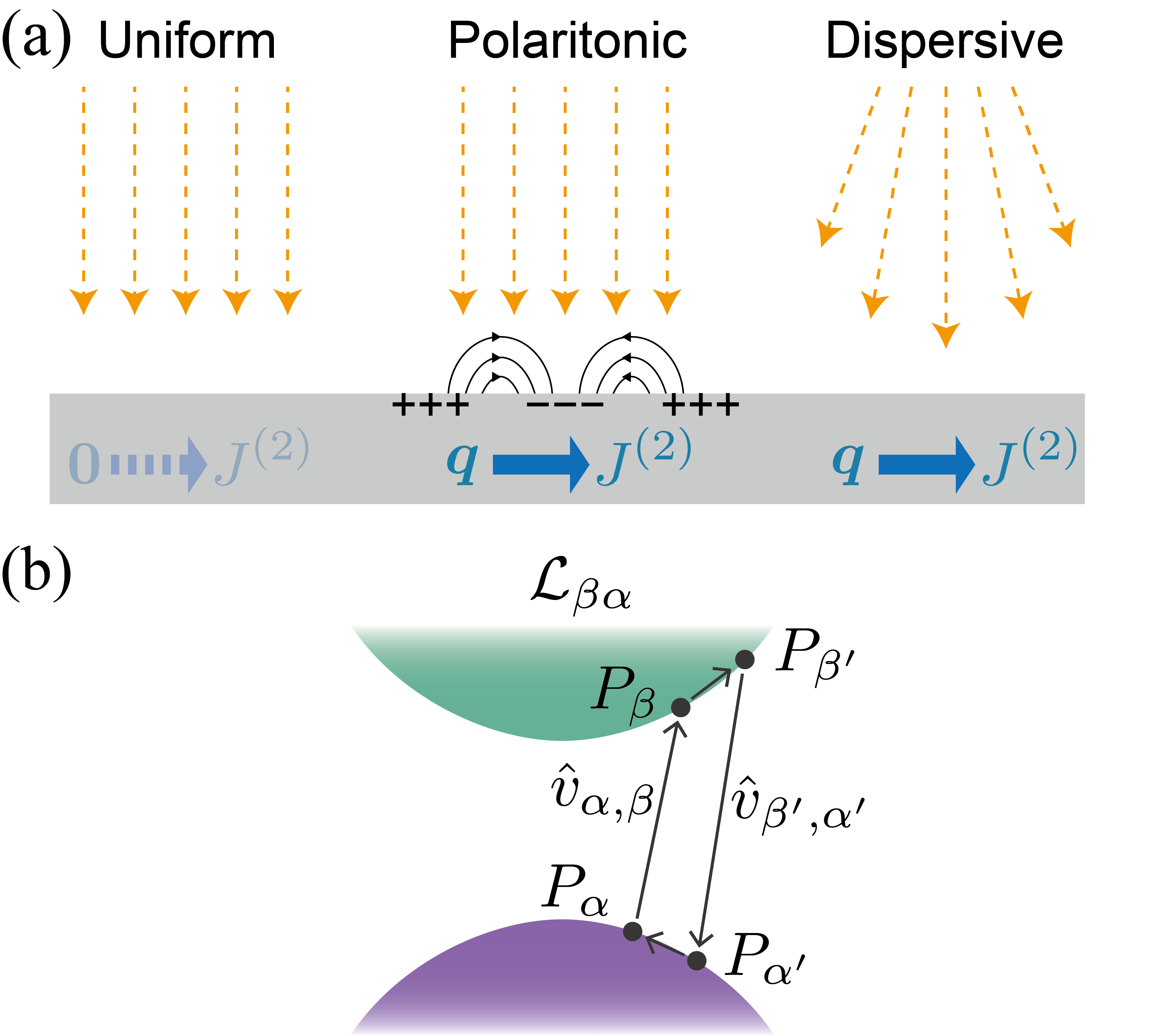}
    \caption{(a) Polaritonic enhancement or spatially dispersive beams can generate a sizable photon-drag effect, circumventing symmetry constraints and producing a nonlinear photon-drag current $J^{(2)}$. (b) Schematic of the geometric loop $\mathcal{L}_{\beta\alpha}$. The loop connects four Bloch states $\alpha, \beta, \alpha'$, and $\beta'$, with $\alpha, \alpha'$ in the valence band and $\beta, \beta'$ in the conduction band. The momenta of these states are related by $\boldsymbol{k}_\beta - \boldsymbol{k}_\alpha = \boldsymbol{k}_{\beta'} - \boldsymbol{k}_{\alpha'} = \boldsymbol{q}$ and $\boldsymbol{k}_{\beta'} - \boldsymbol{k}_\beta = \boldsymbol{k}_{\alpha'} - \boldsymbol{k}_\alpha = \boldsymbol{p}$, where $\boldsymbol{p}$ represents an infinitesimal momentum displacement.}
    \label{fig:schematic}
\end{figure}

\section{Results}\label{Results}
\subsection{Geometric loop formalism}\label{Theory}

We now develop a geometric-loop formulation of the nonlinear photon-drag current, a photovoltaic response in which finite-momentum photoexcitation is essential, as schematized in Fig.~\ref{fig:schematic}a. The inhomogeneous electric field responsible for this effect is given by $\boldsymbol{E}(\boldsymbol{r}, t) = \boldsymbol{E}(\omega) e^{i\boldsymbol{q} \cdot \boldsymbol{r} - i\omega t} + \text{c.c.}$, where $\boldsymbol{q}$ is the photon wavevector and $\omega$ is the frequency. From a density-matrix derivation, the total nonlinear photon-drag current density $J^{a,(2)}$ can be expressed as
\begin{equation}
    J^{a,(2)} = \left( \sigma_{\text{IC}}^{abc} + \sigma_{\text{SC}}^{abc} \right) E_{b}(\omega) E_{c}(-\omega),
\end{equation}
where $\sigma_{\text{IC}}^{abc}$ represents the injection current contribution, while $\sigma_{\text{SC}}^{abc}$ accounts for the shift current contribution (see Supplemental Information for the detailed derivation). For brevity, the $\boldsymbol{q}$ and $\omega$ dependences of $J$ and $\sigma$ are suppressed, and summation over repeated Cartesian indices is implied.

The photon-drag injection-current photoconductivity $\sigma_{\text{IC}}^{abc}$ is expressed as
\begin{equation}
    \sigma_{\text{IC}}^{abc} = 2\tau \mathcal{C} \sum_{n_{\alpha}, n_{\beta}, \boldsymbol{k}} \mathcal{J}_{\alpha\beta}(\omega) \left( v_{\beta \beta}^{a} - v_{\alpha \alpha}^{a} \right) v_{\beta \alpha}^{b} v_{\alpha \beta}^{c},
    \label{eq:pdic}
\end{equation}
where $\tau$ is the carrier relaxation time, $\mathcal{C} = \pi e^{3}/\hbar^{2} \omega^{2}$, and $\sum_{\boldsymbol{k}}$ denotes the continuum limit $\int d^d k / (2\pi)^d$. The composite indices $\alpha = (n_{\alpha}, \boldsymbol{k}_{\alpha})$ and $\beta = (n_{\beta}, \boldsymbol{k}_{\beta})$ refer to Bloch states before and after photoexcitation, characterized by the shifted wavevectors $\boldsymbol{k}_{\alpha} = \boldsymbol{k} - \boldsymbol{q}/2$ and $\boldsymbol{k}_{\beta} = \boldsymbol{k} + \boldsymbol{q}/2$. The term $\mathcal{J}_{\alpha\beta}(\omega) = (f_{\alpha} - f_{\beta})\,\delta(\omega_{\beta} - \omega_{\alpha} - \omega)$ is the joint density of states representing nonvertical transitions of energy $\hbar\omega$ from state $\alpha$ to $\beta$, where $\hbar\omega_\alpha$ and $\hbar\omega_\beta$ denote the respective band energies. The velocity matrix elements are given by $v_{\alpha \beta}^{b} = \langle u_{\alpha} | \hat{v}_{\alpha,\beta}^b | u_{\beta} \rangle$, defined via the midpoint velocity operator $\hat{v}_{\alpha,\beta}^b \equiv \frac{1}{2}(\hat{v}_{\boldsymbol{k}_{\alpha}}^b + \hat{v}_{\boldsymbol{k}_{\beta}}^b)$.

The photon-drag shift-current photoconductivity $\sigma_{\text{SC}}^{abc}$ is given by
\begin{equation}
    \sigma_{\text{SC}}^{abc} = \mathcal{C} \sum_{n_{\alpha}, n_{\beta}, \boldsymbol{k}} \mathcal{J}_{\alpha\beta}(\omega) \left( R_{\beta \alpha}^{b;a} - R_{\alpha \beta}^{c;a} \right) v_{\beta \alpha}^{b} v_{\alpha \beta}^{c},
    \label{eq:pdsc}
\end{equation}
where the shift vector is defined as
\begin{equation}
    R_{\beta \alpha}^{b;a} = i\partial_{k^{a}}\log v_{\beta \alpha}^{b} + r_{\beta \beta}^{a} - r_{\alpha \alpha}^{a},
    \label{eq:sv}
\end{equation}
with $r_{\alpha \alpha}^{a}$ denoting the intra-band Berry connection. Eq.~(\ref{eq:pdsc}) admits a transparent interpretation: it represents the wavepacket shift between the conduction and valence bands, $R_{\beta\alpha}^{b;a}$, weighted by the transition matrix elements $v_{\alpha\beta}^{c}v_{\beta\alpha}^{b}$ and the joint density of states $\mathcal{J}_{\alpha\beta}(\omega)$. Despite the manifest gauge invariance of $R_{\beta\alpha}^{b;a}$, its formal definition relies on the derivatives of gauge-dependent velocity matrices and Berry connections, which severely hinders direct numerical implementation. Alternatively, one can employ the sum-rule expression:
\begin{equation}
    i R_{\beta \alpha}^{b;a} v_{\beta \alpha}^{b} = -w_{\beta \alpha}^{ba} + \sum_{n_\gamma \neq n_\beta} \frac{v_{\beta \gamma}^{a} v_{\gamma \alpha}^{b}}{\omega_{\gamma \beta}} - \sum_{n_\gamma \neq n_\alpha} \frac{v_{\beta \gamma}^{b} v_{\gamma \alpha}^{a}}{\omega_{\alpha \gamma}},
    \label{eq:gdsr}
\end{equation}
where $w_{\alpha \beta}^{ca} = \langle u_{\alpha} | \partial_{k^{a}} \hat{v}_{\alpha, \beta}^{c} | u_{\beta} \rangle$ is the inverse effective mass matrix. Note that to properly account for the finite momentum transfer $\boldsymbol{q}$ in the intermediate states, the wavevectors for the virtual bands inside the summations are implicitly shifted; specifically, $\boldsymbol{k}_{\gamma} = \boldsymbol{k} + \boldsymbol{q}/2$ for the summation over $n_\gamma \neq n_\beta$, and $\boldsymbol{k}_{\gamma} = \boldsymbol{k} - \boldsymbol{q}/2$ for $n_\gamma \neq n_\alpha$. While Eq.~(\ref{eq:gdsr}) facilitates direct numerical evaluation, it obscures the physical intuition by necessitating summations over all energy bands, including those completely uninvolved in the optical transition. Furthermore, converging these infinite sums requires incorporating a vast number of remote bands, imposing a substantial computational burden where truncation inherently compromises numerical precision.

To combine geometric transparency with numerical efficiency, we derive alternative expressions for the shift and injection contributions to the photon-drag response. Before introducing the recast form, we define the geometric loop as
\begin{equation}
    \mathcal{L}_{\beta \alpha}^{abc}(p) = \langle P_{\beta^{\prime}} \hat{v}_{\beta^{\prime}, \alpha^{\prime}}^{b} P_{\alpha^{\prime}} P_{\alpha} \hat{v}_{\alpha, \beta}^{c} P_{\beta} \rangle_{\beta},
    \label{eq:W}
\end{equation}
with $\alpha = (n_{\alpha}, \boldsymbol{k}_{\alpha})$, $\alpha^{\prime} = (n_{\alpha}, \boldsymbol{k}_{\alpha} + p\hat{\boldsymbol{a}})$, $\beta = (n_\beta, \boldsymbol{k}_{\beta})$, and $\beta^{\prime} = (n_\beta, \boldsymbol{k}_{\beta} + p\hat{\boldsymbol{a}})$, where $p$ is an infinitesimal momentum displacement ($p \to 0^+$), $\hat{\boldsymbol{a}}$ is the unit vector along the Cartesian direction $a$, $P_{\alpha} = |u_{\alpha}\rangle\langle u_{\alpha}|$ is the quantum-state projector, and $\langle \dots \rangle_{\alpha}$ denotes the expectation value $\langle u_{\alpha} | \dots | u_{\alpha} \rangle$. Constructed as a gauge-invariant concatenation of quantum-state projectors and velocity operators, $\mathcal{L}_{\beta \alpha}^{abc}(p)$ forms a closed loop that intuitively captures the underlying quantum geometry of the finite-momentum optical transition, as illustrated in Fig.~\ref{fig:schematic}b.

Rewriting the photon-drag shift-current photoconductivity in terms of $\mathcal{L}_{\beta \alpha}^{abc}(p)$ yields:
\begin{equation}
    \sigma_{\text{SC}}^{abc} = \mathcal{C} \sum_{n_{\alpha}, n_{\beta}, \boldsymbol{k}} \mathcal{J}_{\alpha\beta}(\omega) \left( \hat{D}_p \mathcal{L}_{\beta \alpha}^{abc}(p) - \hat{D}_p \mathcal{L}_{\alpha \beta}^{acb}(p) \right),
    \label{eq:pdsc_W}
\end{equation}
where $\hat{D}_p = i \partial_p$ behaves as the generalized dipole operator in parameter space. In this representation, the velocity product and shift vector from Eq.~(\ref{eq:pdsc}) are compactly expressed as $v_{\beta \alpha}^{b} v_{\alpha \beta}^{c} = \mathcal{L}_{\beta \alpha}^{abc}(p)$ and $R_{\beta \alpha}^{b;a} = \hat{D}_p \log \mathcal{L}_{\beta \alpha}^{abc}(p)$. Eq.~(\ref{eq:pdsc_W}) indicates that the quantum-geometric content of $\sigma_{\text{SC}}^{abc}$ is fully encoded in the dipole of the geometric loop. Because $\mathcal{L}_{\beta \alpha}^{abc}(p)$ is manifestly gauge invariant, its derivative can be robustly evaluated numerically via finite differences. Using the same strategy, the photon-drag injection-current photoconductivity becomes
\begin{equation}
    \sigma_{\text{IC}}^{abc} = 2\tau \mathcal{C} \sum_{n_{\alpha}, n_{\beta}, \boldsymbol{k}} \mathcal{J}_{\alpha\beta}(\omega) \left( v_{\beta \beta}^{a} - v_{\alpha \alpha}^{a} \right) \mathcal{L}_{\beta \alpha}^{abc}(p).
    \label{eq:pdic_W}
\end{equation}
In this formulation, the quantum-geometric origin of $\sigma_{\text{IC}}^{abc}$ is captured directly by $\mathcal{L}_{\beta \alpha}^{abc}(p)$, weighted by the group velocity difference between the conduction and valence bands. 

Physically, the geometric loop $\mathcal{L}_{\beta \alpha}^{abc}(p)$ represents a generalized quantum holonomy tailored to describe nonlinear finite-momentum transitions. Unlike traditional quantum geometric quantities defined locally at a single momentum $\boldsymbol{k}$, $\mathcal{L}_{\beta \alpha}^{abc}(p)$ constructs a closed, bi-local path in reciprocal space—connecting distinct shifts $\boldsymbol{k}_\alpha$ and $\boldsymbol{k}_\beta$ across the valence and conduction bands. The complex quantum geometry encoded in this closed path directly dictates the real-space displacement of the wavepacket (extracted as the shift vector via the logarithmic derivative) without resorting to artificially fragmented sum rules. Consequently, the geometric loop formalism emerges as the fundamental building block for modeling nonlinear photon-drag responses in realistic material systems.

\subsection{Symmetry perspective}\label{Symmetry}
Spatial and time-reversal symmetries play a pivotal role in governing the electronic properties of quantum materials. Here, we derive how these symmetries constrain the photoconductivity tensors $\sigma_{\text{IC}}^{abc} $ and $\sigma_{\text{SC}}^{abc}$. We begin by expanding the response in powers of the photon wavevector:
\begin{equation}
\sigma^{abc}(\boldsymbol{q})=\sigma^{abc}+q_{d}\,\sigma^{abcd}+q_{d}q_{e}\,\sigma^{abcde}+\cdots,
\label{eq:Taylor}
\end{equation}
where the expansion coefficients $\sigma^{abc}, \sigma^{abcd}, \sigma^{abcde}, \dots$ are independent of $\boldsymbol{q}$, and summation over repeated spatial indices is implied. Because conductivity tensors of even and odd ranks transform differently under spatial operations, and noting from Eq.~(\ref{eq:Taylor}) that odd-rank (even-rank) tensors couple exclusively with even (odd) powers of $\boldsymbol{q}$, we decompose the total response into components that are even and odd in $\boldsymbol{q}$:
\begin{equation}
\sigma^{abc}(\boldsymbol{q})=\sigma^{abc}_{\text{even}}(\boldsymbol{q})+\sigma^{abc}_{\text{odd}}(\boldsymbol{q}).
\label{eq:even_odd}
\end{equation}
In the perturbative regime of small $\boldsymbol{q}$, $\sigma^{abc}_{\text{even}}(\boldsymbol{q})$ is dominated by the conventional bulk photovoltaic contribution ($\boldsymbol{q}^0$), whereas the leading contributions to $\sigma^{abc}_{\text{odd}}(\boldsymbol{q})$ are the first-order ($\boldsymbol{q}^1$) injection-current and shift-current dipoles. 

The symmetry-enforced constraints on $\sigma_{\text{IC}}^{abc}$ and $\sigma_{\text{SC}}^{abc}$ under a spatial operation $\mathcal{G}$, time-reversal symmetry $\mathcal{T}$, and their combination $\mathcal{GT}$ are summarized in Table~\ref{t01}. Within this unified framework, $\mathcal{G}$ denotes the defining spatial symmetry of the system: while $\mathcal{G}=\mathcal{P}$ applies to the full tensor ($a,b,c \in \{x,y,z\}$) and a three-dimensional wavevector $\boldsymbol{q}$ in bulk materials, $\mathcal{G}=C_{2z}$ acts as the spatial operation governing the in-plane responses ($a,b,c \in \{x,y\}$) and an in-plane wavevector $\boldsymbol{q}$ for 2D architectures such as pristine TBG. A comprehensive symmetry derivation for the nonlinear photon-drag photoconductivity tensor is provided in the Supplemental Information.

\begin{table}[t] 
\centering
\caption{\textbf{Symmetry analysis of allowed photoconductivity components.} Allowed components of $\sigma_{\text{IC}}^{abc}$ and $\sigma_{\text{SC}}^{abc}$ are listed under various symmetry operations, where $\mathcal{G}$ represents either inversion $\mathcal{P}$ or $C_{2z}$ rotation, and $\mathcal{T}$ denotes time-reversal symmetry. While $\mathcal{G}=\mathcal{P}$ imposes constraints on all tensor components, $\mathcal{G} = C_{2z}$ restricts only the in-plane components. Here, $\sigma_{\text{even}}^{abc}$ and $\sigma_{\text{odd}}^{abc}$ denote response tensors that are even and odd functions of the photon wavevector $\boldsymbol{q}$, respectively. Labels L and C identify the linear (real) and circular (imaginary) parts that remain invariant under the symmetry; their absence signifies that the component must vanish. The symbol $\times$ marks configurations where both parts are strictly forbidden.}
\label{t01}
\addtolength{\tabcolsep}{4pt} 
\renewcommand{\arraystretch}{1.3} 
\begin{tabular}{ccccc}
\toprule
Symmetry & $\sigma^{abc}_{\text{IC},\text{even}}(\boldsymbol{q})$ & $\sigma^{abc}_{\text{IC},\text{odd}}(\boldsymbol{q})$ & $\sigma^{abc}_{\text{SC},\text{even}}(\boldsymbol{q})$ & $\sigma^{abc}_{\text{SC},\text{odd}}(\boldsymbol{q})$ \\
\midrule
$\mathcal{G}$ & $\times$ & L, C & $\times$ & L, C \\
$\mathcal{T}$ & C & L & L & C \\
$\mathcal{GT}$ & L & L & C & C \\
$\mathcal{G}$ \& $\mathcal{T}$ & $\times$ & L & $\times$ & C \\
\bottomrule
\end{tabular}
\end{table}

\subsection{Twisted bilayer graphene}\label{TBG1}
Motivated by the moiré enhancement of photon-drag effects, we use TBG as a case study. We employ an exact continuum model based on the ab initio informed framework of Carr et al.~\cite{carr2019}. Relative to the well-known Bistritzer-MacDonald model~\cite{bistritzer2011}, this model incorporates three essential improvements: (i) atomic relaxation taken from elastic-theory-relaxed geometries, including in-plane strain and out-of-plane corrugation; (ii) moiré couplings beyond the first shell, necessary to capture stacking-order variations at small angles; and (iii) explicit $\boldsymbol{k}$-dependent terms that enable the $\boldsymbol{k}\cdot\boldsymbol{p}$ model to reproduce more accurately the particle-hole asymmetry of realistic ab initio bands. The combination of fidelity and efficiency makes this continuum model a practical backbone for our investigation of the nonlinear photon-drag currents in TBG. Although pristine TBG is noncentrosymmetric, its $D_6$ point group contains a $C_{2z}$ rotation that maps $(x, y) \to (-x, -y)$ and prohibits all in-plane photoconductivity components of the bulk photovoltaic effects, effectively mirroring the constraints of inversion for these components. To investigate the nonlinear photon-drag currents in TBG, we focus exclusively on these in-plane components $\sigma^{abc}(\boldsymbol{q})$ ($a, b, c \in \{x, y\}$) in the following discussion.

\begin{figure}[htbp]
    \centering
    \includegraphics[width=1.0\columnwidth]{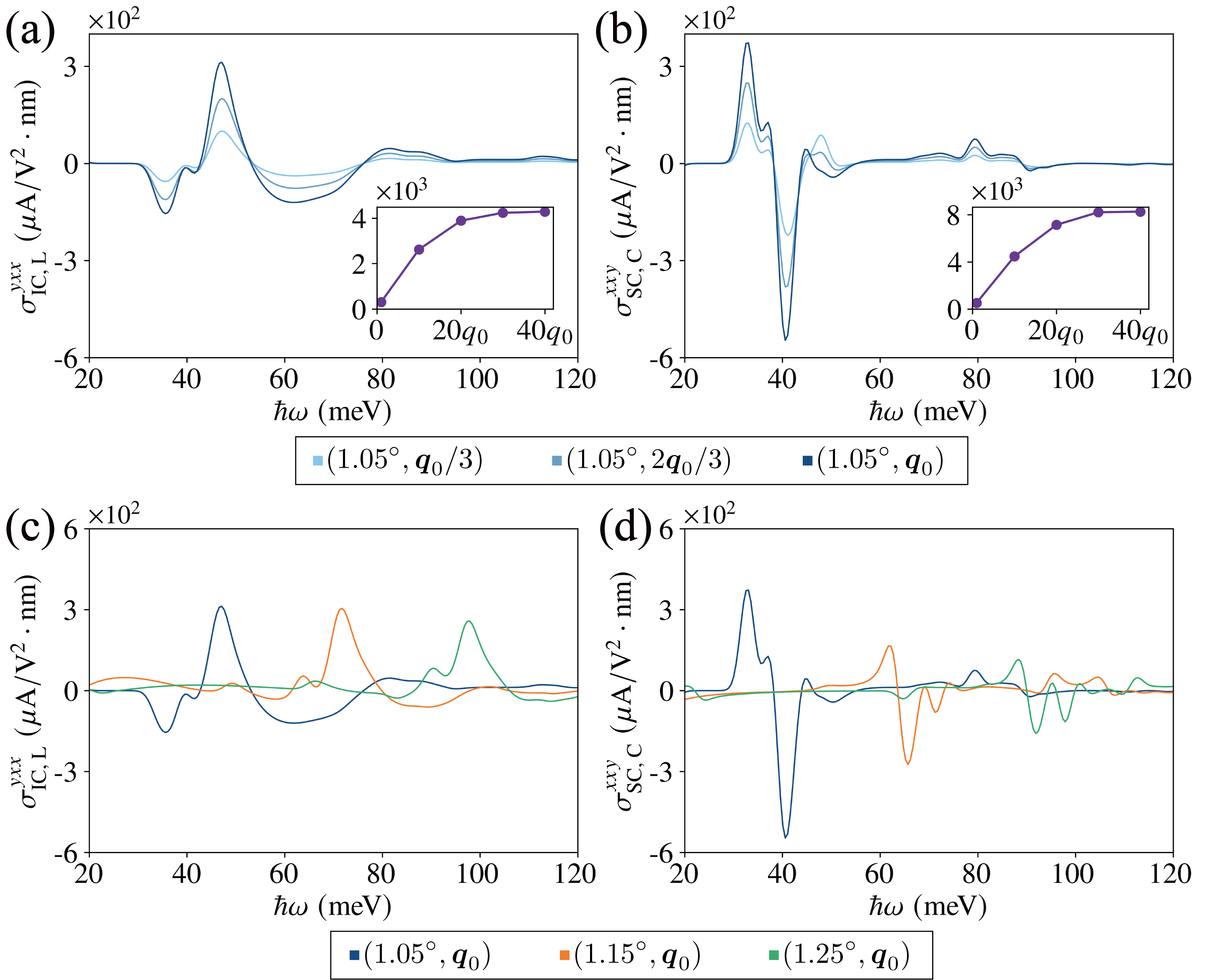}
    \caption{Photon-drag photoconductivity of pristine TBG. 
    (a) Moiré lattice at a small twist angle. The two layers form a hexagonal moiré pattern with high-symmetry stacking regions.
    (b) Band structures along $\rm K\!-\!\Gamma\!-\!M\!-\!K$ at twist angles $\theta=1.05^\circ$, $1.15^\circ$, and $1.25^\circ$. The bands are maximally flat near $\theta=1.05^\circ$ and broaden progressively with increasing $\theta$.
    (c) $\sigma_{\text{IC,L}}^{yxx}$ and (d) $\sigma_{\text{SC,C}}^{xxy}$ spectra at $\theta=1.05^\circ$, showing the evolution with the photon wavevector; curves from light to dark correspond to $\boldsymbol{q}_0/3$, $2\boldsymbol{q}_0/3$, and $\boldsymbol{q}_0$. For (c), we use a moderate relaxation time $\tau=10^{-13}~\mathrm{s}$. 
    (e--f) Spectra at a fixed wavevector $\boldsymbol{q}=\boldsymbol{q}_0$ for twist angles $\theta=1.05^\circ$, $1.15^\circ$, and $1.25^\circ$, respectively. 
    (g--h) Magnitude of the peak photoconductivity versus $\boldsymbol{q}$ over an extended range, evaluated at the spectral peaks $\hbar\omega = 47~\text{meV}$ for (c) and $\hbar\omega = 41~\text{meV}$ for (d).}
    \label{fig:TBG1}
\end{figure}

We begin by examining the numerical results obtained from pristine TBG. The underlying moiré lattice at small twist angles is illustrated in Fig.~\ref{fig:TBG1}a. The low-energy band structures for three twist angles ($\theta=1.05^\circ$, $1.15^\circ$, and $1.25^\circ$) are plotted along the high-symmetry path in Fig.~\ref{fig:TBG1}b. At $\theta=1.05^\circ$, which is slightly above the magic angle, the bands exhibit their maximum flatness. As $\theta$ increases, the moiré bandwidths broaden progressively, shifting spectral features to higher energies relative to the charge neutrality point. Unless otherwise specified, we set the photon wavevector to $\boldsymbol{q}_0=(0,10^{-4}~\text{\AA}^{-1})$ for all photon-drag calculations. According to the selection rules derived in Table~\ref{t01}, only the $\boldsymbol{q}$-odd linear injection and circular shift currents contribute to the response. Among these configurations, $C_{2x}$ symmetry dictates that only components featuring an odd number of $y$ indices remain non-vanishing. 

Figs.~\ref{fig:TBG1}c and \ref{fig:TBG1}d show the calculated spectra at $\theta=1.05^\circ$ as $\boldsymbol{q}$ varies from $\boldsymbol{q}_0/3$ to $2\boldsymbol{q}_0/3$ and $\boldsymbol{q}_0$. Within this small-$\boldsymbol{q}$ window, the peak magnitudes of $\sigma^{yxx}_{\text{IC,L,odd}}(\boldsymbol{q})$ and $\sigma^{xxy}_{\text{SC,C,odd}}(\boldsymbol{q})$ grow approximately linearly with $\boldsymbol{q}$, indicating that the response is dominated by the first-order injection- and shift-current dipoles. Figs.~\ref{fig:TBG1}e and \ref{fig:TBG1}f compare the spectra across the three twist angles; the dominant peaks systematically shift to higher photon energies as $\theta$ increases, directly tracking the bandwidth broadening. Figs.~\ref{fig:TBG1}g and \ref{fig:TBG1}h quantify the scaling behavior over an extended range of $\boldsymbol{q}$, showing that the peak values increase by nearly an order of magnitude before saturating. Alongside this momentum-dependent saturation, the response curves reveal a pronounced twist-angle dependence. Both the injection and shift mechanisms of the nonlinear photon-drag currents weaken at larger twist angles because the expanding moiré Brillouin zone effectively diminishes the relative influence of the fixed photon momentum $\boldsymbol{q}$. Beyond this macroscopic suppression, however, their decay rates diverge at the microscopic level. The circular shift current collapses precipitously as its underlying interband quantum geometry simplifies. For optical excitations in the $20\text{--}120~\text{meV}$ range, the response is entirely dominated by transitions between the moiré flat bands and the dispersive bands. Near the magic angle, these transitions connect extremely localized and highly itinerant states, entailing a drastic spatial reconstruction that creates a giant mismatch between the wavepacket centers and yields a large shift vector. With increasing twist angle, bandwidth broadening delocalizes the flat-band electrons, diminishing this spatial discrepancy and driving the sharp decay of the shift response. Conversely, the linear injection current decays much more gradually: band steepening at larger angles enhances the group velocity difference, providing a kinetic compensation that partially offsets the overall decay. 

To gauge the absolute scale of these effects, we take the pronounced peak in $\sigma^{xxy}_{\text{SC,C,odd}}(\boldsymbol{q})$ near $\hbar\omega\simeq 41~\text{meV}$ at $\theta=1.05^\circ$ as a reference point. A free-space photon at this energy possesses a wavevector of $q=\omega/c\approx 2.03\times 10^{-5}~\text{\AA}^{-1}$. Thus, a modest in-plane wavevector only about five times larger already yields substantial nonlinear photocurrents, which are an order of magnitude larger than the response predicted for the prototypical ferroelectric GeS monolayer~\cite{rangel2017} and highly comparable to the giant responses predicted for magnetic axion insulators such as $\text{MnBi}_2\text{Te}_4$ bilayers~\cite{wang2020, fei2020, xu2021}.

\begin{figure}[htbp]
    \centering
    \includegraphics[width=1.0\columnwidth]{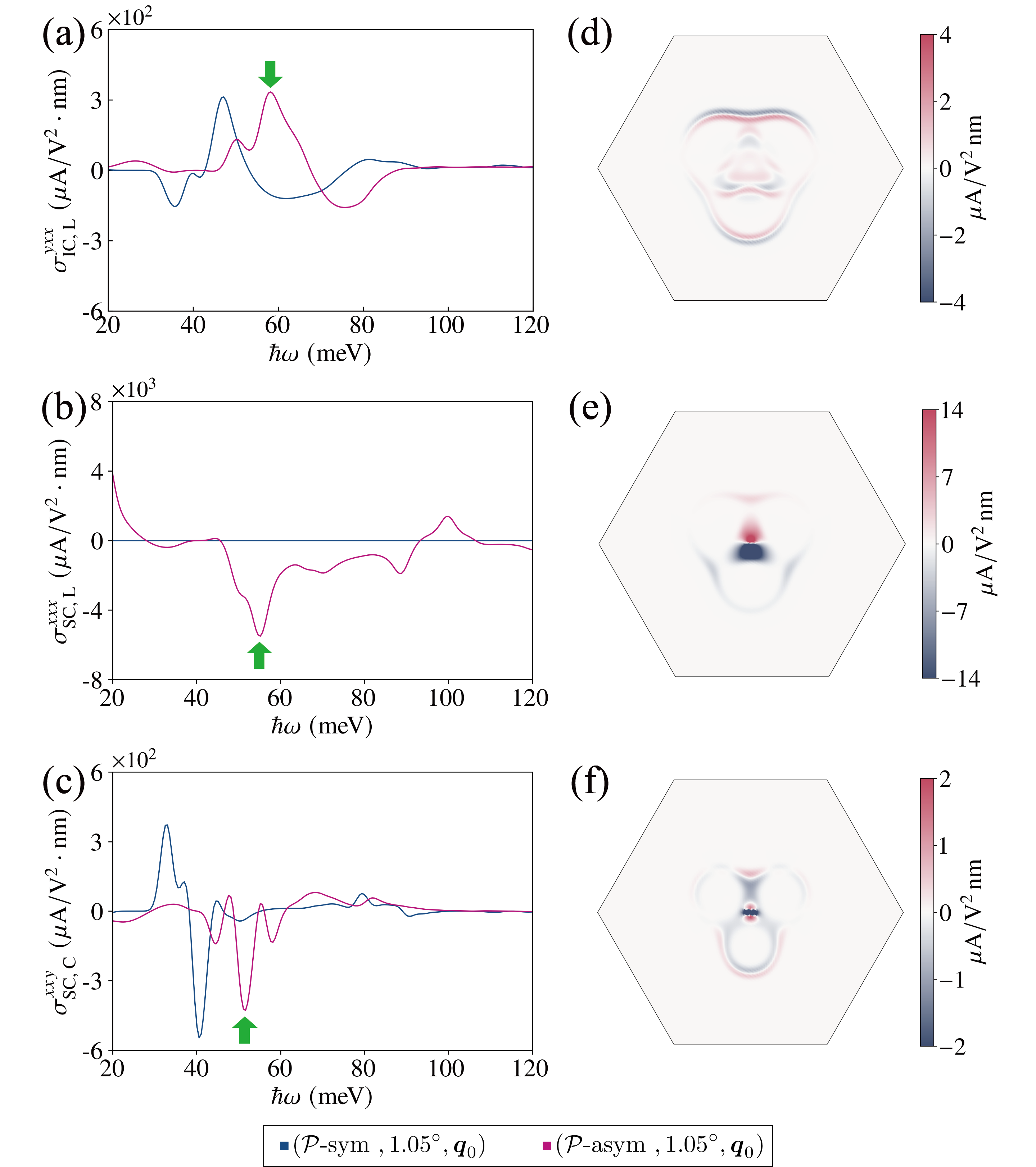}
    \caption{Influence of symmetry breaking on photon-drag photoconductivity in TBG.  
    (a) Schematic of the $D_3$-symmetric moiré lattice, where a sublattice-staggered potential $\pm\delta_S$ distinguishes the $A$ and $B$ sites. 
    (b) Comparison of moiré band structures at $\theta = 1.05^\circ$ for $D_6$ (blue) and $D_3$ (purple) configurations. 
    (c--e) Photoconductivity spectra comparing $D_6$-symmetric and $D_3$-symmetric TBG at $\theta = 1.05^\circ$.
    (f--h) $\boldsymbol{k}$-resolved photoconductivity corresponding to (c--e) evaluated at the specific photon energies indicated by the green arrows: (f) $\hbar\omega=58~\text{meV}$, (g) $\hbar\omega=55~\text{meV}$, and (h) $\hbar\omega=51~\text{meV}$.}
    \label{fig:TBG2}
\end{figure}

To further evaluate the relative magnitudes of the nonlinear photon-drag and conventional bulk photovoltaic currents in TBG, we incorporate the influence of a hexagonal boron nitride (hBN) substrate. This substrate lowers the point-group symmetry from $D_{6}$ to $D_{3}$, thereby lifting the spatial symmetry constraints that otherwise forbid the conventional in-plane bulk photovoltaic effect. We model this environment as a sublattice-staggered potential, $\pm\delta_{S}$, setting $\delta_{S}=0.02~\text{eV}$ on both layers for simplicity~\cite{carr2019a}. The resulting $D_3$-symmetric moiré lattice is illustrated in Fig.~\ref{fig:TBG2}a, where the staggered potential effectively distinguishes the $A$ and $B$ sites within each layer. This symmetry reduction exerts a profound impact on the low-energy electronic structure. As shown in the comparative band structure in Fig.~\ref{fig:TBG2}b, the introduction of $\pm\delta_S$ lifts the original Dirac point degeneracy at the $K$ point of the mini-Brillouin zone. Specifically, at the twist angle $\theta = 1.05^\circ$, the pristine flat bands are modified by the opening of a distinct single-particle band gap, driving a semimetal-to-semiconductor transition. 

Under $D_3$ symmetry, the conventional contribution ($\boldsymbol{q}^0$) to $\sigma_{\text{IC,C,even}}^{abc}$ is strictly forbidden by its antisymmetry under $b \leftrightarrow c$ field exchange, whereas the corresponding contribution to $\sigma_{\text{SC,L,even}}^{abc}$ remains allowed due to its symmetric nature. Furthermore, under $C_{2x}$ symmetry, only the specific components of $\sigma_{\text{SC,L,even}}^{abc}$ with an even number of $y$ indices remain non-vanishing. Consistent with this symmetry deduction, our numerical simulations demonstrate that $\sigma_{\text{SC,L,even}}^{xxx}(\boldsymbol{q})$ vanishes identically in pristine TBG but emerges prominently when $\delta_{S} \neq 0$, as shown in Fig.~\ref{fig:TBG2}d. The $\boldsymbol{q}$-odd responses, namely $\sigma_{\text{IC,L,odd}}^{yxx}(\boldsymbol{q})$ and $\sigma_{\text{SC,C,odd}}^{xxy}(\boldsymbol{q})$, exhibit modest line-shape modifications that closely track the gap opening driven by the broken symmetry, as shown in Figs.~\ref{fig:TBG2}c and \ref{fig:TBG2}e. 

Comparing Fig.~\ref{fig:TBG2}d with Figs.~\ref{fig:TBG2}c and \ref{fig:TBG2}e, we observe that the conventional linear shift current exceeds the linear injection and circular shift currents by roughly an order of magnitude, directly reflecting the small-$\boldsymbol{q}$ scaling behavior. However, once $\boldsymbol{q}$ reaches the saturation regime, the $\boldsymbol{q}$-odd photon-drag response is expected to become comparable to the $\boldsymbol{q}$-even conventional response. To clarify the microscopic origin of these signals, we evaluate the $\boldsymbol{k}$-resolved photoconductivity at representative spectral peaks. As shown in Figs.~\ref{fig:TBG2}f--\ref{fig:TBG2}h, the finite response is entirely dominated by interband transitions occurring near the $\Gamma$ point, specifically from the flat valence band to the dispersive conduction band. In $\boldsymbol{k}$-space, the textures of $\sigma_{\text{IC,L,odd}}^{yxx}(\boldsymbol{q})$ and $\sigma_{\text{SC,C,odd}}^{xxy}(\boldsymbol{q})$ display alternating positive-negative lobes arranged in closely spaced parallel stripes, whereas $\sigma_{\text{SC,L,even}}^{xxx}(\boldsymbol{q})$ completely lacks this oscillating feature. This clear contrast follows from the former two responses being governed by the injection- and shift-current dipoles, while the latter arises purely from the conventional shift current.

\section{Conclusion}\label{Conclusion}

In summary, we have developed a geometric-loop formalism that unifies the photon-drag injection and shift-current contributions within a single, transparent framework, providing clear quantum-geometric insights while remaining numerically efficient. Within this framework, we have derived general macroscopic selection rules and symmetry constraints on the nonlinear photon-drag photoconductivity tensor. When applied to TBG using a realistic continuum model that incorporates atomic relaxation and particle-hole asymmetry, our formalism reveals that even modest in-plane photon wavevectors can induce substantial nonlinear photocurrents in pristine TBG. Near twist angles slightly above the magic value ($\approx 1.05^\circ$), the predicted photon-drag injection and shift currents reach magnitudes that rival the giant bulk photovoltaic responses reported in prototypical 2D semiconductors. The scaling behavior with photon momentum is approximately linear in the small-$\boldsymbol{q}$ regime, serving as a clear signature of the underlying injection- and shift-current dipole mechanisms. As the photon wavevector increases further, the peak response grows by nearly an order of magnitude before eventually saturating. We have traced the evolution of these photon-drag spectra, demonstrating that the prominent spectral peaks systematically shift to higher photon energies as the moiré bandwidth broadens with increasing twist angle. Finally, for TBG aligned with an hBN substrate, the resulting sublattice asymmetry breaks the structural $C_{2z}$ symmetry and activates the conventional bulk photovoltaic effect. This conventional shift current ($\boldsymbol{q}^0$) dominates in the small-$\boldsymbol{q}$ regime, exceeding the leading $\boldsymbol{q}^1$ photon-drag contributions by roughly an order of magnitude. Crucially, the $\boldsymbol{k}$-space textures of the conventional shift current differ qualitatively from those of the photon-drag contributions, underscoring their distinct microscopic origins. Overall, these results demonstrate that exploiting photon momentum provides a potent knob for generating massive nonlinear dc photocurrents in moiré architectures, opening up promising avenues for photon-drag-engineered optoelectronic functionalities.

\section{Methods}\label{Methods}

\textbf{TBG Hamiltonian.} We describe the low-energy electronic structure of TBG using an ab initio-informed $\boldsymbol{k}\cdot\boldsymbol{p}$ continuum model that incorporates the crucial role of lattice relaxation and achieves the accuracy of full density functional theory calculations \cite{fang2016c, fang2018, carr2018, carr2019}. The underlying $\boldsymbol{k}\cdot\boldsymbol{p}$ parameters are directly extracted from a microscopic ab initio tight-binding Hamiltonian constructed for supercells spanning the twist-angle range $0.18^\circ \le \theta \le 6^\circ$. These parameters exhibit a smooth dependence on $\theta$, enabling precise interpolation between specific configurations to generate an effective model valid for any continuous angle within this domain. The total moiré Hamiltonian $H_{K}$ for the $K$ valley is formulated as:
\begin{equation}  
H_{K}=\left[\begin{array}{cc}
H_D^1(\boldsymbol{k})+A_1(\boldsymbol{r})+V_1(\boldsymbol{r}) & \tilde{T}^{\dagger}(\boldsymbol{r})+\left\{M_{\pm}^{\dagger}(\boldsymbol{r}), \hat{k}_{-}\right\}+\left\{M_{-}^{\dagger}(\boldsymbol{r}), \hat{k}_{-}\right\} \\
\tilde{T}(\boldsymbol{r})+\left\{M_{+}(\boldsymbol{r}), \hat{k}_{+}\right\}+\left\{M_{-}(\boldsymbol{r}), \hat{k}_{-}\right\} & H_D^2(\boldsymbol{k})+A_2(\boldsymbol{r})+V_2(\boldsymbol{r})
\end{array}\right]
\end{equation}  
The diagonal blocks represent the intralayer Hamiltonians, where $H_{D}^{i}(\boldsymbol{k})$ is the isolated Dirac Hamiltonian for layer $i$. The term $A_{i}(\boldsymbol{r})$ denotes the in-plane pseudogauge field arising from internal lattice deformation and strain, which is expanded in Fourier components as $A_i(\boldsymbol{r})=\sum_{\boldsymbol{p}_i} A_{\boldsymbol{p}_i}^i e^{i \boldsymbol{p}_i \cdot \boldsymbol{r}}$, where $\boldsymbol{p}_i=n \boldsymbol{G}_1+m \boldsymbol{G}_2$ for integers $m, n$ and moiré supercell reciprocal lattice vectors $\boldsymbol{G}_i$. The potential $V_{i}(\boldsymbol{r})$ accounts for external site-energy modulations, such as the sublattice-staggered potential induced by an hBN substrate or electrical gating. The off-diagonal blocks describe the generalized interlayer tunneling. Departing from the standard rigid-lattice Bistritzer-MacDonald model, this framework employs a generalized interlayer scattering term $\tilde{T}(\boldsymbol{r}) = \sum_{\boldsymbol{q}_i} \tilde{T}_{\boldsymbol{q}_i} e^{i \boldsymbol{q}_i \cdot \boldsymbol{r}}$. In this expansion, the structural momentum transfer $\boldsymbol{q}_i$ (which is purely a lattice property and distinct from the photon wavevector $\boldsymbol{q}$) is defined as $K_1 - K_2 + \boldsymbol{G}$, where $K_i$ represents the original Dirac point of the Brillouin zone for layer $i$ and $\boldsymbol{G}$ denotes a reciprocal lattice vector of the moiré supercell. This generalized formulation incorporates higher-order Fourier components essential for accurately capturing the intricate variations in stacking order that emerge upon relaxation at small twist angles. Furthermore, the model includes momentum-dependent scattering terms $M_{\pm}(\boldsymbol{r})$ (where $\hat{k}_{\pm} = \hat{k}_{x} \pm i\hat{k}_{y}$), which are crucial for properly reproducing the particle-hole asymmetry observed in realistic ab initio band structures. By fully accounting for both in-plane and out-of-plane atomic relaxations, the model provides a high-fidelity description of the moiré flat bands. Notably, this relaxation shifts the primary magic angle to $\theta_c \approx 1.0^\circ$ and eliminates the spurious second magic angle inherent to unrelaxed models.

The Hamiltonian defined above corresponds to the $K$ valley. Given that the $K$ and $K'$ valleys are related by time-reversal symmetry $\mathcal{T}$, the total photoconductivity tensor is obtained by summing the contributions from both valleys:
\begin{equation}
\sigma^{abc}_{\text{total}}(\boldsymbol{q})
=\sigma^{abc}_{\text{cmp}, K}(\boldsymbol{q}) \pm \sigma^{abc}_{\text{cmp}, K}(-\boldsymbol{q}),
\label{eq:PD_total}
\end{equation}
where the plus sign applies to the $\boldsymbol{q}$-even components ($\text{cmp} = \text{IC,C}$ or $\text{SC,L}$) and the minus sign applies to the $\boldsymbol{q}$-odd components ($\text{cmp} = \text{IC,L}$ or $\text{SC,C}$). A detailed symmetry-based derivation mapping the $K'$ valley response directly to the $K$ valley via Eq.~(\ref{eq:PD_total}) is provided in the Supplemental Information.

\section*{Acknowledgements}
H.W. acknowledges the support from the NSFC under Grants Nos. 12522411, 12474240, and 12304049, as well as the support from Fundamental Researeh Funds for the Central Universities. K. C. acknowledges the support from the NSFC under Grants No. 12488101, and the Innovation Program for Quantum Science and Technology under Grant No. 2024ZD0300104.

% \backmatter


% \bibliography{ref}% common bib file

\end{document}